\documentclass [12pt]{article}
\usepackage{amsmath,amssymb}
\usepackage[dvips]{graphicx}
\newcommand{\bs}{\begin{subequations}}
\newcommand{\es}{\end{subequations}}

\numberwithin{equation}{section}

\def \myfigures #1#2#3#4#5#6#7#8
{\begin{figure}[ht]
      \centering\includegraphics[width=#2 \textwidth]{#1.eps}
        \hfill
      \centering\includegraphics[width=#6 \textwidth]{#5.eps}
        \parbox[t]{#4\textwidth}{\caption {#3}\label{#1}}
        \hfill
        \parbox[t]{#8\textwidth}{\caption {#7}\label{#5}}
\end{figure} }

\begin{document}

\title{Static vortices in long Josephson junctions of exponentially varying width}

\author{E.G. Semerdzhieva\thanks{Plovdiv University, Bulgaria; E-mail: elis@jinr.ru}, 
T.L. Boyadzhiev\thanks{Joint Institute for Nuclear Research, Dubna 141980, Russia; Sofia University, Bulgaria; E-mail: todorlb@jinr.ru}, 
Yu. M. Shukrinov\thanks{Joint Institute for Nuclear Research, Dubna 141980, Russia; Polytechnical Institute, Dushanbe, Tajikistan; E-mail: shukrinv@thsun1.jinr.ru}}

\date{}
\maketitle

\begin{center}
    Submitted January 16, 2004, \\ Low Temperature Physics (Russian) \textbf{30}, 610--618 (June 2004)
\end{center}

\begin{abstract}

\pagenumbering{arabic}\thispagestyle{myheadings}

A numerical simulation is carried out for static vortices in a long
Josephson junction with an exponentially varying width. At specified
values of the parameters the corresponding boundary-value problem
admits more than one solution. Each solution (distribution of the
magnetic flux in the junction) is associated to a Sturm--Liouville
problem, the smallest eigenvalue of which can be used, in a
first approxi\-mation, to assess the stability of the vortex against relatively small
spatiotemporal perturbations. The change in width of the junction leads
to a renormalization of the magnetic flux in comparison with the case
of a linear one-dimensional model. The influence of the model
parameters on the stability of the states of the magnetic flux is
investigated in detail, particularly that of the shape parameter. The
critical curve of the junction is constructed from pieces of the
critical curves for the different magnetic flux distributions having the
highest critical currents for the given magnetic field.

\end{abstract}

\section{Statement of the problem}

\begin{figure}[ht]
    \centering\includegraphics[totalheight=5cm,keepaspectratio]{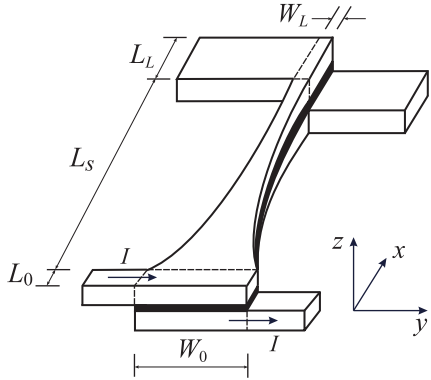}
        \caption{Diagram of the Josephson junction of exponentially varying
width.} \label{model1}
\end{figure}

A Josephson junction model that takes into account the influence of the
shape in the $xy$ plane was considered in Refs.\ \cite{1,2,3}.
For a model junction with generatrices that vary by an exponential
law \cite{2,3} (see Fig.\ \ref{model1}) the basic equation for the phase
$\varphi (t,x)$ in the junction can be written in the form
\begin{equation} \label{time}
    \ddot \varphi + \alpha \dot\varphi - \varphi\,'' + \sin \varphi + g(t,x) = 0\,, \quad x \in (0,l)\,,\quad t>0\,.\\
\end{equation}
The spatial coordinate $x$ is normalized to the Josephson penetration
depth $\lambda_{J}$, and $l = (L_0+L_{s}+L_{L})/\lambda_{J}$. The
time $t$ is normalized to the plasma frequency (see, e.g., the
monograph cited as Ref.\ \cite{4}). The parameter $\alpha $
describes dissipative effects in the junction. An overdot denotes
differentiation with respect to time $t$, and a prime denotes
differentiation with respect to the spatial coordinate $x$.

The term $g(t,x)\equiv \sigma [\varphi\,'(t,x)-h_{B}]$ in Eq.\ \eqref{time}
models the current caused by the exponential variation of the width of
the junction. The dimensionless shape parameter $\sigma  \geq 0$ is
determined by the following expression (see Fig.\ \ref{model1} for
notation):
\begin{eqnarray*}
 \sigma = \frac{\lambda_J}{L_s}\,\ln \left(\frac{W_0}{W_l}\right)\,,
\end{eqnarray*}
where it is assumed that $W_0>W_{L}$. It is known that after suitable
normalization the function $\varphi (t,x)$ can be interpreted as the
(dimensionless) magnetic flux along the junction. Then the value of
$h_{B}$ is the (dimensionless) strength of the external (boundary)
magnetic field along the $y$ axis. For simplicity it is assumed below
that the field $h_{B}$ is independent of time.

A detailed derivation of Eq.\ \eqref{time} is given in Ref.\ \cite{2}.

In the present paper we restrict consideration to a model junction with
an in-line geometry. If the injection current density through the end
$L_0$ of the junction is denoted by $\gamma $ (which is assumed
constant), the boundary conditions at $x = 0$ and $x = l$, after the
transition to the corresponding limits, take the form
\begin{equation}\label{bc}
    \varphi\,'(t,0) = h_B - l\,\gamma, \quad \varphi\,'(t,l) = h_B\,.
\end{equation}

Because of the presence of a dissipative term $\alpha \dot{\varphi }$
the solution $\varphi (t,x)$ of equation \eqref{time} can lose energy and,
consequently, for $t\rightarrow \infty $ it will approach the
corresponding static distribution $\varphi_{s}(x)$ (to simplify the
notation below we shall use the subscript $s$ when necessary). This
makes it necessary to investigate in detail the static distributions in
the Josephson junction and their behavior upon variations of the model
parameters.

The boundary-value problem for the static distributions follows
directly from relations \eqref{time} and \eqref{bc}:
\bs\label{stat}
    \begin{gather}
        -\varphi\,'' + \sin \varphi  + g(x) = 0\,,\quad x\in(0,l)\,, \label{steq}\\
        \varphi\,'\left( 0 \right) - h_B  + l\gamma  = 0, \label{bca} \\
        \varphi\,'\left( l \right) - h_B  = 0\,, \label{bcb}
    \end{gather}
\es
where the ``geometric'' current is given by the expression
\begin{equation} \label{geom}
    g(x) \equiv \sigma \left[ \varphi\,'(x) - h_B \right]\,.
\end{equation}
We note that the solutions of the problem \eqref{stat} depends not only on the
spatial coordinate $x$ but also on the set of parameters $p\equiv
\{l,\sigma ,h_{B},\gamma \}$, i.e., $\varphi(x,p)$. When necessary below
we shall indicate the dependence of the quantities on the parameters.

Let $\varphi_{s}(x)$ be some static solution of equation \eqref{time}, i.e., a solution of the boundary-value problem \eqref{stat}. With the goal of investigating the stability of the solution $\varphi_{s}(x)$, following Ref.\ \cite{5}, we consider a spatiotemporal fluctuation of the form
\begin{equation}\label{fluct}
    \varphi(t,x) = \varphi_s (x) + \varepsilon\, e^{-\alpha\,t/2} \sum\limits_n {\left[ {e^{i\omega_n t} \psi_n(x) + e^{-i\omega_n t} \psi_n^*(x)} \right]}\,,
\end{equation}
which depends on the small parameter $\varepsilon $. Substituting the
expansion \eqref{fluct} into Eq.\ \eqref{time} and conditions \eqref{bc}, to a first approximation
in $\varepsilon $ we arrive at the Sturm--Liouville (SL) problem
\bs\label{c3}
    \begin{gather}
         -\psi\,'' + \sigma \psi\,' + q\,(x)\, \psi  = \lambda \,\psi \,,\quad x\in(0,l)\,, \label{slp} \\
        \psi\,'(0) = 0, \quad \psi\,'(l) = 0\,,
    \end{gather}
\es
the potential of which, $q(x)\equiv \cos\varphi_{s}(x)$, is determined
by the known static solution $\varphi_{s}(x)$. Here $\lambda = (\omega ^{2}+\alpha ^{2}/4)^{1/2}$ is the spectral parameter.

In view of the boundedness of the function $|q(x,p)| \leq 1$ on a
finite interval of the variable $x$ there exists\cite{6} a countable
sequence, bounded from below, of real eigenvalues $\lambda_{\rm
min}\equiv \lambda_0<\lambda_{1}<\lambda_{2}<\ldots <\lambda_{n}<\ldots $ of problem \eqref{c3}. To each eigenvalue $\lambda_{n}$ ($n 0,1,2,\ldots )$ there corresponds a single real eigenfunction $\psi_{n}(x)$ satisfying the normalization condition
\begin{equation}\label{norm}
    \int\limits_0^l{\psi^2_n(x)\,dx} = 1\,.
\end{equation}
LHere the number of zeroes of the eigenfunction $\psi_{n}(x)$ on the
interval $(0,l)$ is equal to the index $n$. In particular, the
eigenfunction $\psi_0(x)$ corresponding to the smallest eigenvalue
$\lambda_{\rm min}$ does not have zeroes on $(0,l)$.

Since $\varphi (x,p)$ depends on the set of parameters $p$, the
potential of the SL problem and, hence, the corresponding eigenvalues
and eigenfunctions of the SL problem also depend on the parameters,
i.e., $\lambda_{n} = \lambda_{n}(p)$ and $\psi_{n} = \psi
_{n}(x,p)$.

We shall say \cite{5} that in a certain range of variation of the
parameters ${\cal P}\subset \mathbb{R}^{4}$ the static solution
$\varphi_{s}(x)$ is stable in a first approximation with respect to
spatiotemporal perturbations of the form \eqref{fluct} if in that region $\lambda
_{\rm min}(p)>0$. If $\lambda_{\rm min}(p)<0$, then a component that
is rapidly growing in time appears in the expansion \eqref{fluct}, and the
solution $\varphi_{s}(x)$ is unstable. The points of the vector of
parameters which lie on the hypersurface
\begin{equation}\label{biff}
    \lambda_{min}(p) = 0
\end{equation}
in the space ${\cal P}$ are points of bifurcation (branching) for the
solution $\varphi (x,p)$. The values of the parameters for which Eq.\
\eqref{biff} holds are called the bifurcation or critical values for the
solution $\varphi (x)$. The corresponding bifurcation curves for the
remaining two parameters are given by the sections of the surface \eqref{biff}
by hyperplanes corresponding to fixed values of the two parameters. The
most important from the standpoint of the possibility of experimental
verification are the critical curves of the current versus magnetic
field type,
\begin{equation}\label{gamhb}
    \lambda_{min}(\gamma, h_B) = 0\,,
\end{equation}
corresponding to fixed geometric parameters $l$ and $\sigma $ of the junction.

From a theoretical standpoint knowledge of the bifurcation curves
enables one to find the number of equilibrium solutions, to understand
their structure, and to describe the physics of the phenomenon.
Numerical simulation simplifies the study and makes it possible to
estimate the range of variation of the parameters in which one can
expect stability or instability of the magnetic flux distributions in
the Josephson junction.

Especially important for practical purposes is the ability to check
experimentally the bifurcation curves of the parameters of the
Josephson junction, which in turn is an important source of information
for refining the physical model. As a concrete example let us indicate
the methods of studying soliton-like vortex structures of the magnetic
flux in Josephson junctions on the basis of measurements of the
magnetic-field dependence of the critical current (see, e.g., Refs.\
\cite{3,5} and \cite{7,8,9,10}).

We note that a number of papers have been devoted to the laying of a
rigorous mathematical groundwork for reducing the problem of stability
of the solutions of nonlinear operator equations to an investigation of
eigenvalue problems for a linear operator (see, e.g., the
review \cite{11} and the collected papers \cite{12} and the literature
cited therein).

Josephson junctions with an exponentially varying width in the $xy$
plane have been studied theoretically and experimentally in Refs.\
\cite{2} and \cite{3}. The influence of the shape on the
current--voltage characteristics of the junctions was studied in detail.
However, the problems involving the determination of the stability regions
of the static distributions and the structure of the critical curves
have not been adequately studied. The present paper is devoted to a
study of those questions, which are important from the applied and
theoretical points of view.

\section{Numerical results and discussion}

The exact analytical solutions of equation \eqref{steq} for the case $\sigma = 0$ are expressed in terms of elliptic functions \cite{13}. For $\sigma >0$ approximate methods can be used \cite{2}. In both cases a stability analysis of the solutions using the SL problem \eqref{c3}  is difficult in view of the complexity of the corresponding expressions for the potential $q(x)$. Therefore it is advisable to carry out a detailed analysis of the possible states of the magnetic flux in the junction and analysis of their stability with the help of a numerical simulation.

In this paper we use a continuous analog of Newton's method for solution of the boundary-value problems \eqref{stat} and \eqref{c3}, \eqref{norm} (see reviews \cite{14, 15}). The linearized boundary-value problems arising at each iteration were solved numerically with the use of a spline collocation scheme\cite{16} of improved accuracy.

To permit comparison of our results with the results of Ref.\ \cite{3}, the majority of the numerical simulations were carried out for Josephson junctions of length $l = 10$ and $l = 20$.

For an ``infinite'' one-dimensional and uniform Josephson junction ($\sigma  = 0$, $l\rightarrow \infty $, $x\in(-\infty ,\infty)$) the well-known exact analytical expression (see, e.g., Refs.\ \cite{4} and \cite{5})
\begin{equation} \label{1fluxon}
    \varphi (x) = 4\arctan {e^{\, \pm \,x} }\,,
\end{equation}
usually called the fluxon and antifluxon, respectively, is valid. For
realistic Josephson junctions of finite length those entities are
deformed by the geometry of the junction and also by the influence of
the magnetic field $h_{B}$ and/or the external current $\gamma $ and
are not fluxons in the strict sense of the word (the functions \eqref{1fluxon} do
not satisfy conditions \eqref{bc}). However, because of a number of features
of those soliton-like solutions, in particular, their finite energy and
size, it is appropriate and convenient to use these conventional names.
For brevity below we denote the fluxon in the Josephson junction as
$\Phi ^{1}$.

According to \cite{5}, on the entire axis ($-\infty ,\infty
)$ the discrete spectrum of the SL problem generated by solution \eqref{1fluxon}
consists of an isolated point $\lambda  = 0$, i.e., the
fluxon/antifluxon in an ``infinite'' Josephson junction is found in a
quasi-stable (bifurcation) state.

It follows from general comparison theorems for SL problems that for
a finite Josephson junction the condition $\lambda_{\rm min}<0$ holds,
i.e., the stability of $\Phi ^{1}$ becomes worse.

For $\gamma  = 0$ and small $|h_{B}|$ the only stable state in a
Josephson junction of finite length is the Meissner (vacuum)
state, which will be denoted by $M$. For $h_{B} = 0$, $\gamma = 0$ this ``trivial'' solution of problem \eqref{stat}, of the form $\varphi
(x) = 0,\pm 2\pi ,\pm 4\pi ,\ldots $ (there is no magnetic field
in the junction). For these same parameters the smallest
eigenvalue of the SL problem for the $M$ distribution is $\lambda
_{\rm min}(M) = 1$. In addition to $M$ there is also an unstable
Meissner solution $\varphi (x) = \pm \pi, \pm 3\pi, \ldots $, for
which $\lambda_{\rm min} = -1$.
\myfigures{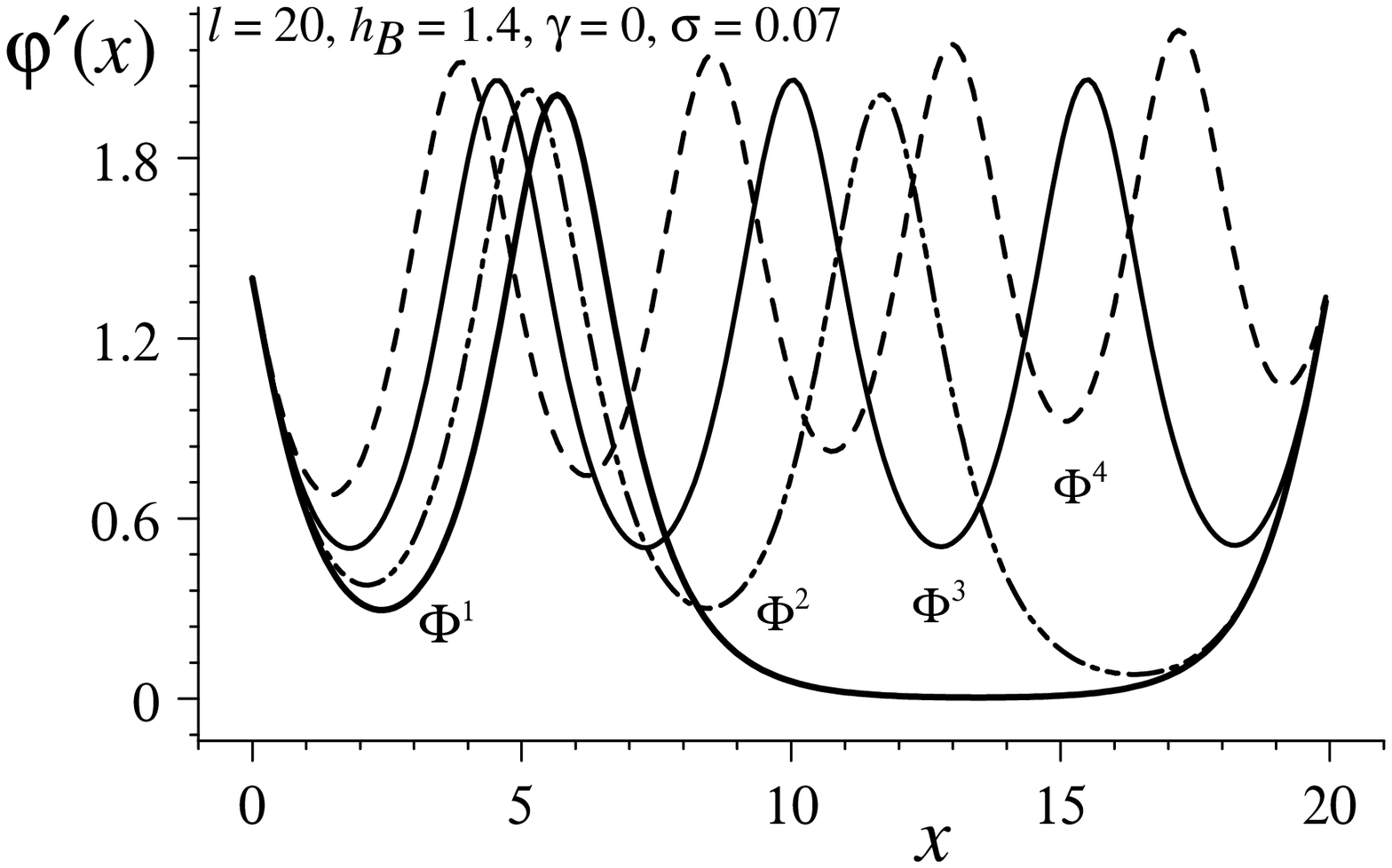}{0.47}{Fluxon vortices in a Josephson junction.}{0.48} {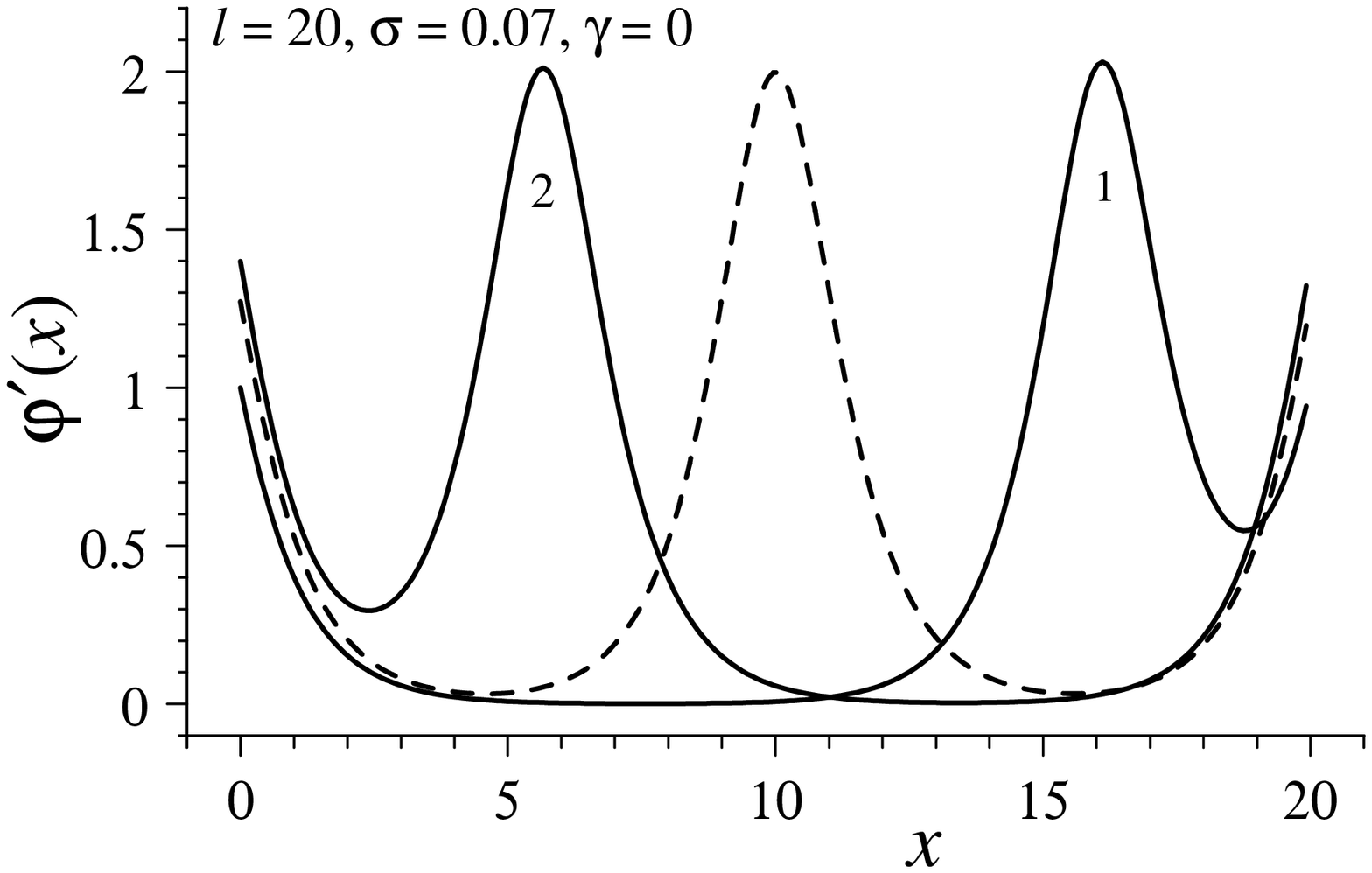}{0.47}{The $\varphi\,'(x)$ curves for different $h_{B}$. }{0.49}

For sufficiently large values of the boundary magnetic field $h_{B}$,
stable fluxon vortices are generated in the Josephson junction. For
example, for $h_{B} = 1.4$, $\gamma  = 0$ and $\sigma  = 0.07$, the Josephson
junction contains, in addition to the $M$ distribution, the multifluxon vortices
$\Phi ^{n}$, $n = 1,2,3,4$, graphs of which are shown in Fig.\ \ref{allsol_hb14}.
The number of vortices is determined by the value of the total magnetic
flux\cite{5} in the Josephson junction:
$$\Delta\varphi = \varphi(l) - \varphi(0)\,.$$

For an ``infinite'' junction (for $l\rightarrow \infty )$ the value
$\Delta \varphi /2\pi \rightarrow k$, where $k = 0,2,3,\ldots$ is the
number of vortices (fluxons) in the distribution $\varphi (x)$. For the
solution in Fig.\ \ref{allsol_hb14} we have, respectively, $\Delta \varphi (\Phi
^{1})/2\pi \approx 1.49$, $\Delta \varphi (\Phi ^{2})/2\pi \approx
2.48$, $\Delta \varphi (\Phi ^{3})/2\pi \approx 3.45$, and $\Delta
\varphi (\Phi ^{4})/2\pi \approx 4.36$, while for the Meissner solution
$\Delta \varphi (M)/2\pi \approx 0.49$.

The influence of the external magnetic field $h_{B}$ on the magnetic
flux distribution $\varphi '(x)$ in the junction for the main fluxon
$\Phi ^{1}$ at $\sigma  = 0.07$ is demonstrated in Fig.\ \ref{dphi_hb}. At a
certain value $h_{B} = h_{m}$ the maximum of the derivative $\varphi
'(x)$ is localized in the middle of the junction (curve {\em 2},
$h_{B}\approx 1.273$). For $h_{B}<h_{m}$ the fluxon is ``expelled'' to
the end $x = l$ by the ``geometric'' current $g(x)$ (curve {\em 1},
$h_{B} = 1$). For $h_{B}>h_{m}$ the fluxon is shifted by the external
field toward the end $x = 0$ (curve {\em 2}, $h_{B} = 1.4)$.

If the length of the Josephson junction is sufficiently large,
then a change of the current $\gamma $, equivalent to a change in
magnetic field at the left end $x = 0$, turns out to have only a
slight influence on the local maximum magnetic field in the
junction, as is well demonstrated by Fig.\ \ref{dfg_hb12}. For
comparison, the situation in which the shape factor $\sigma  = 0$
is demonstrated in Fig.\ \ref{dphi_s0} for the same values of the
remaining parameters. It is seen that variations of the current
cause the maximum of the magnetic field to shift to the right or
left of the center of the Josephson junction, depending on the
direction of the current.\cite{19}
\myfigures{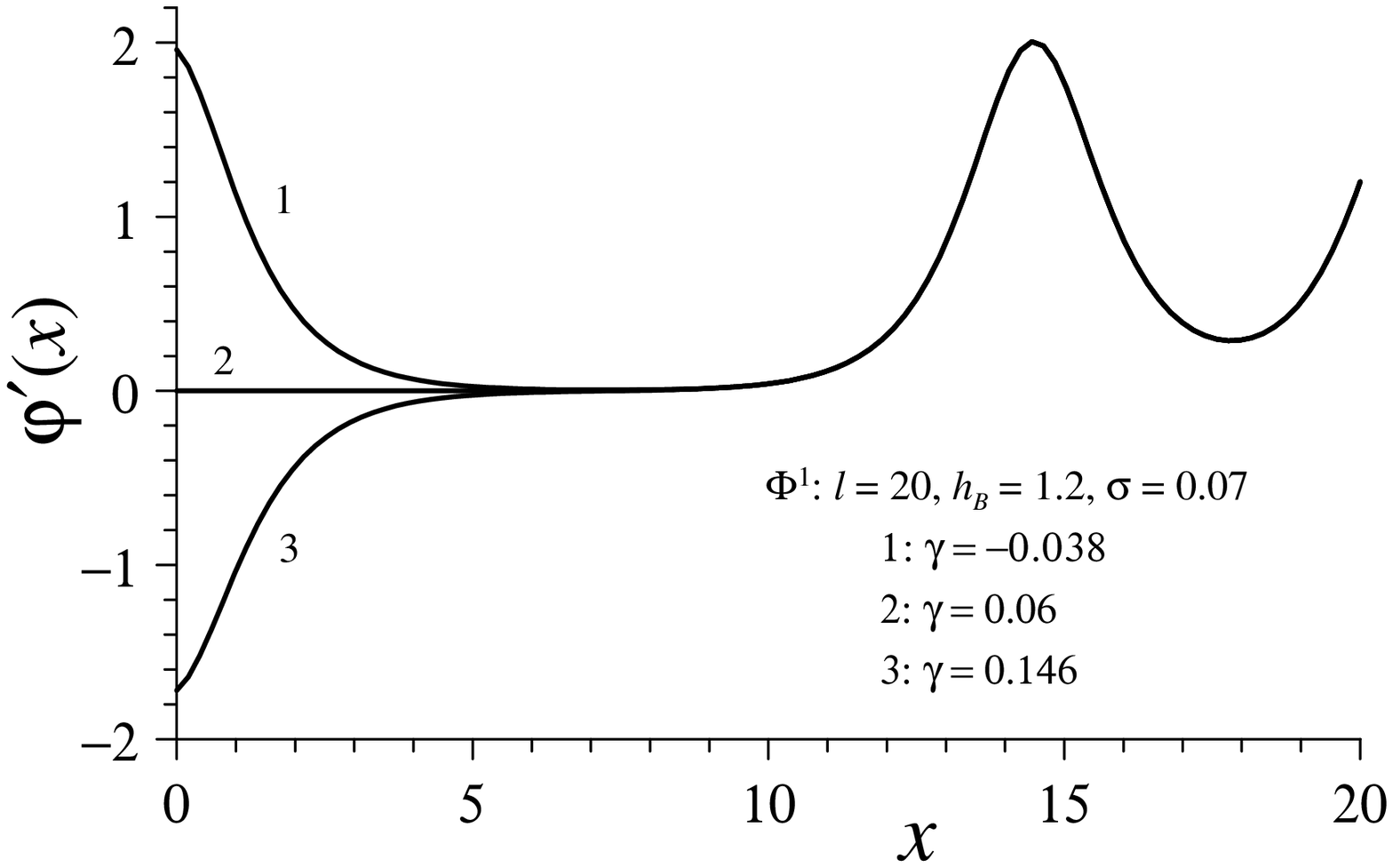}{0.44}{The $\varphi\,'(x)$ curve for $\sigma  = 0.07$.}{0.5} {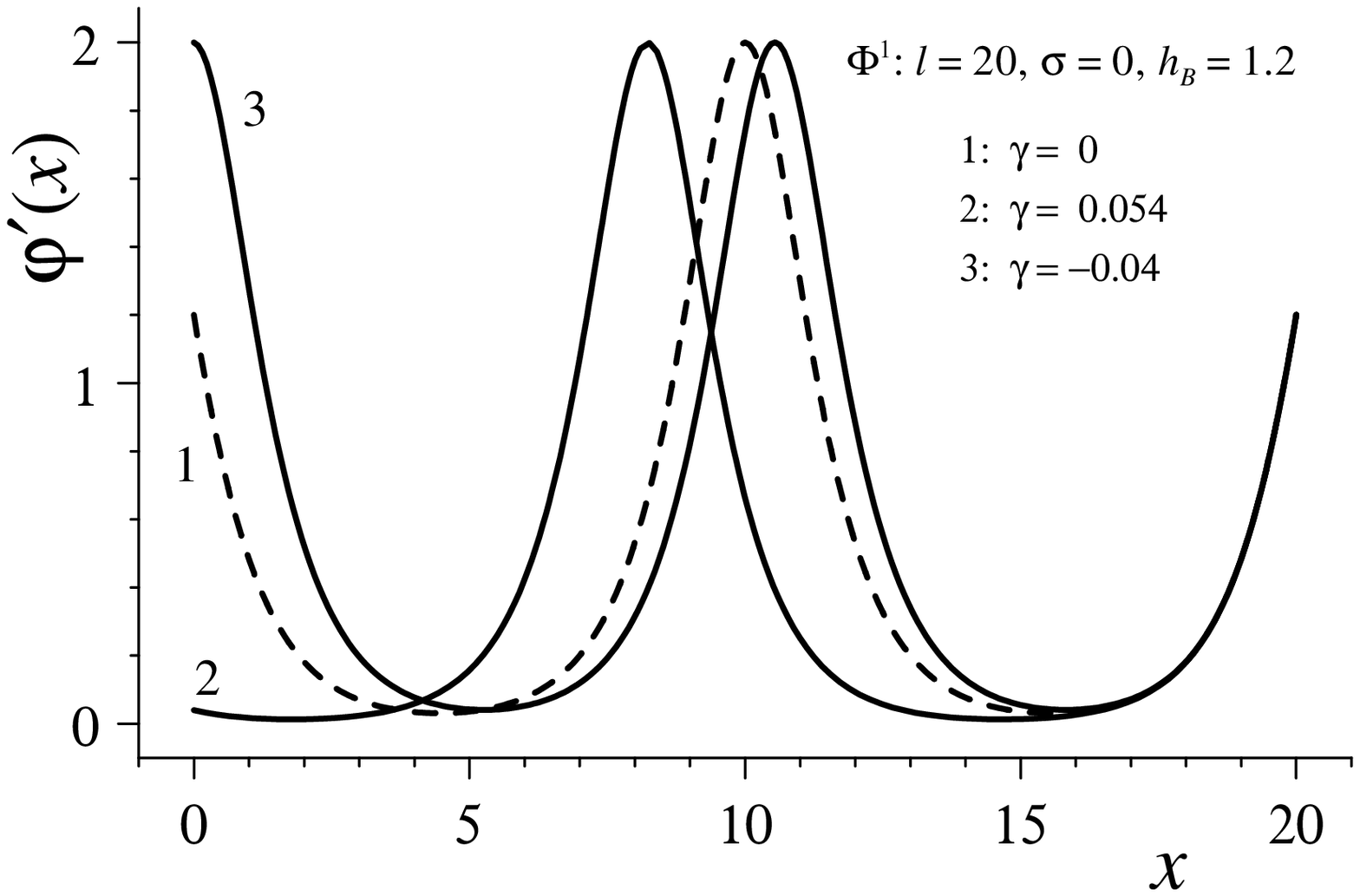}{0.43}{The $\varphi\,'(x)$ curve for $\sigma = 0$.}{0.49}

For values $h_{B}>h_{m}$ the scheme $x_{m}$ of the maximum of the field
$\varphi '(x)$ of the fluxon $\Phi ^{1}$ is shifted to the left from
the center $x = l/2$ toward the $x = 0$ end (see Fig.\ \ref{dphi_hb}). The
``motion'' of the maximum of the function $\varphi '(x)$ upon
variations of the parameters of the problem occurs in accordance with
the equation
\[ \sin \varphi(x_m ,p) + \sigma \left[ {\varphi\,'(x_m, p) - h_B} \right] = 0\,, \]
which expresses the balance of the Josephson and ``geometric'' currents
at the point $x_{m}$. Here for the $\Phi ^{1}$ distribution the
coordinate $x_{m} = l/2$ for $\sigma  = 0$, $\gamma  = 0$, and any
attainable $h_{B}$. This case corresponds to the dashed line in Fig.\
\ref{xmax_hbg0}, which shows graphs of the function $x_{m}(h_{B},\sigma )$ for
several values of the shape parameter $\sigma $. Each curve for $\sigma
>0$ intersects the straight line $x_{m}(h_{B},0) = l/2$ at a certain
point $h_{m}(\sigma ) = \varphi '(l/2,\sigma )$ at which the maximum
magnetic field inside the Josephson junction is centered. It is
important to note that for $\sigma >0$ the $x_{m}(h_{B})$ curves change
sharply in the vicinity of $h_{m}$: slight deviations of the external
field $h_{B}$ from the value $h_{m}$ cause a significant displacement
of the field maximum from the center of the junction.

Let us consider the influence of two geometric parameters of the model,
viz., the shape parameter $\sigma $ and length $l$, on the magnetic flux
distribution in the Josephson junction.

The dependence of the smallest eigenvalue of the SL problem for the
main fluxon $\Phi ^{1}$ on the shape parameter $\sigma $ is shown in
Fig.\ \ref{la_shape}. It is seen that for fixed $h_{B}$ and $\gamma $ there
exists a certain maximum value of $\sigma $ above which the
distribution of $\Phi ^{1}$ loses stability, i.e., a bifurcation of the
vortex occurs upon a change in $\sigma $. Large values of the magnetic
field $h_{B}$ correspond to large critical values of $\sigma $. The
value of the current is important at small values of $\sigma $ and
plays practically no role at values of $\sigma $ close to the maximum.
Figure \ref{la_shape} thereby demonstrates the destructive character of the
shape parameter at large values of it and also the stabilizing role of
the boundary magnetic field $h_{B}$.

\myfigures{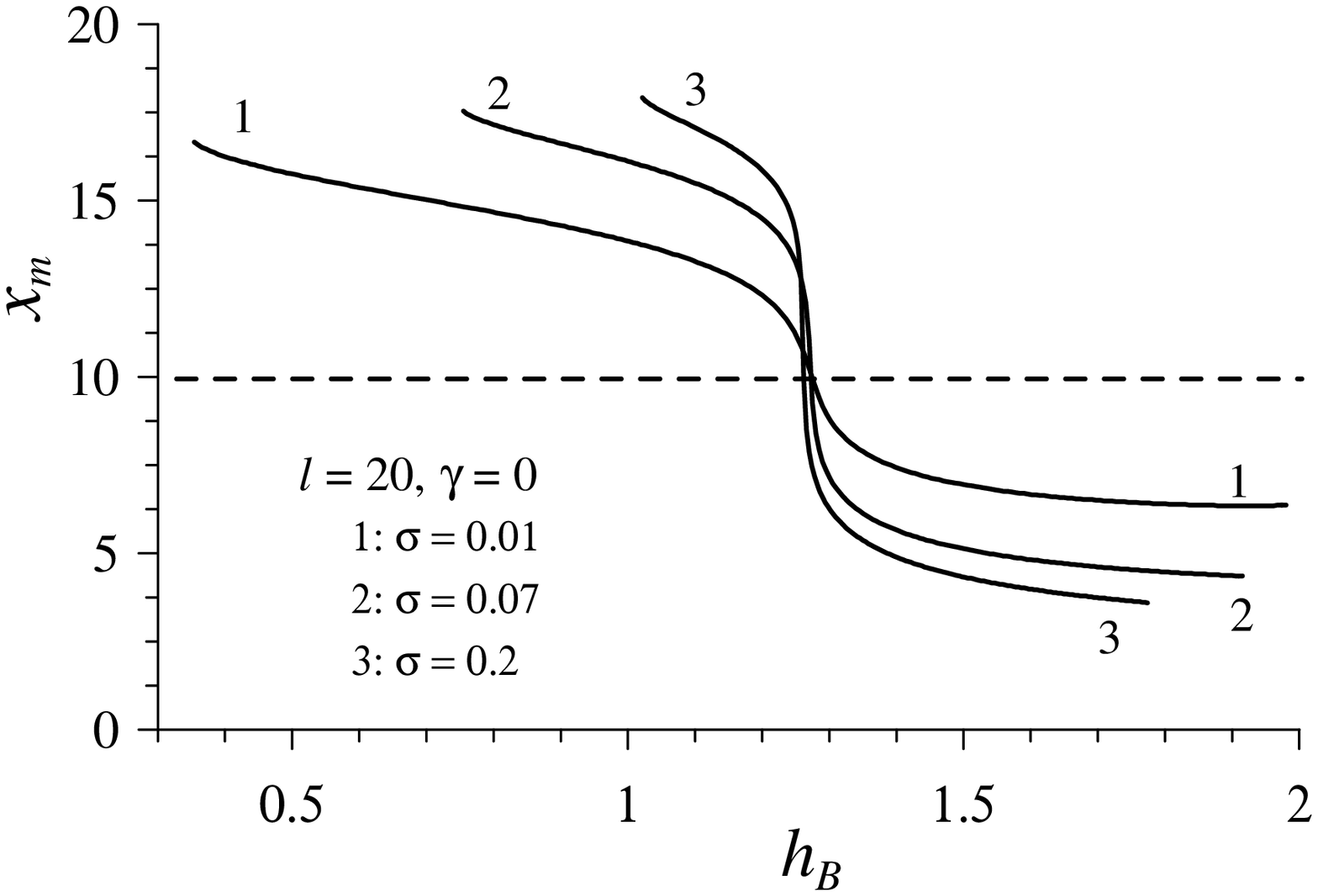}{0.43}{``Movement'' of the maximum of the magnetic field in a
Josephson junction upon a change in $h_{B}$. }{0.5}{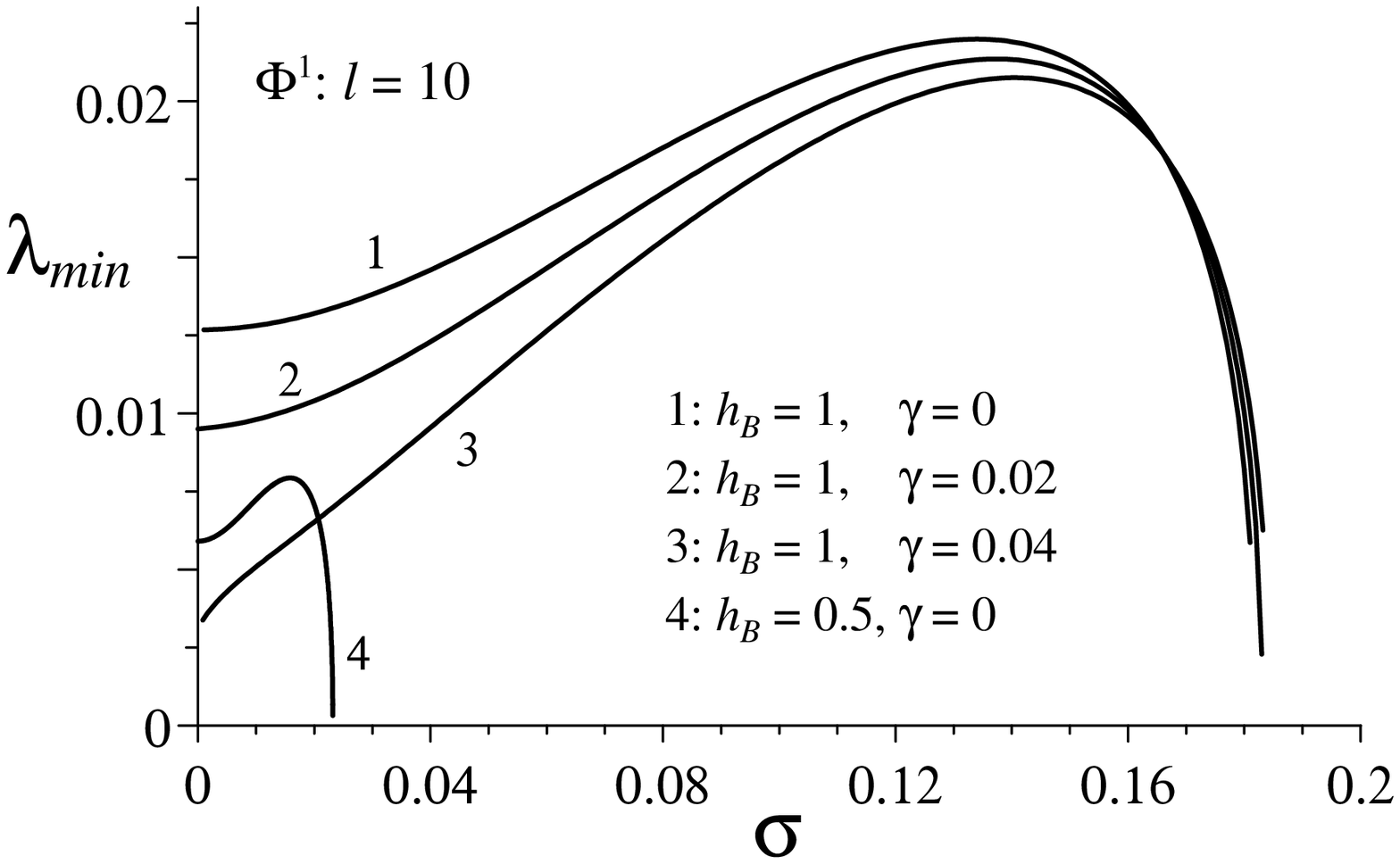}{0.48}{Bifurcation upon a change in $\sigma $. }{0.47}

\begin{figure}[ht]
    \begin{center}
        \includegraphics[totalheight=4.9cm,keepaspectratio]{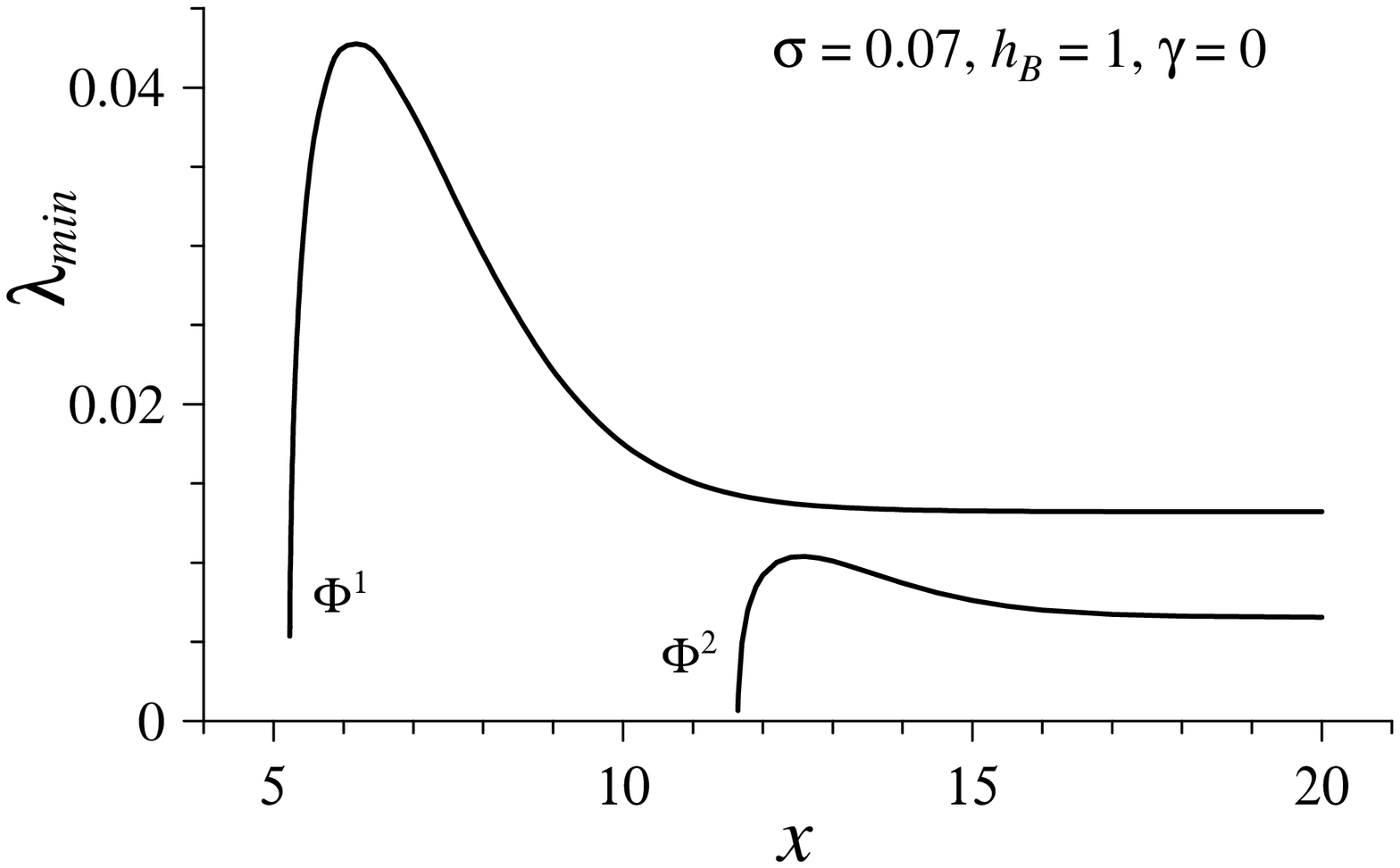}{\caption {Influence of the length of the Josephson junction on thestability of $\Phi ^{1}$ and $\Phi ^{2}$. } \label{lamin_ls07}}
    \end{center}
\end{figure}

The influence of the length $l$ of the Josephson junction on the
stability of the main fluxon $\Phi ^{1}$ is shown in Fig.\ \ref{lamin_ls07}.
It is seen that $\lambda_{\rm min}(l)\approx {\rm const}$ at $l>12$,
and therefore to a certain accuracy the Josephson junction can be
considered ``infinite'' for $\Phi ^{1}$. At $l<6$ the smallest
eigenvalue of the SL problem \eqref{c3} falls rapidly, going to zero at
$l\approx 5.23$. Thus there exists a minimum length of the junction for
which the fluxon $\Phi ^{1}$ maintains stability.\cite{18} The minimum
length clearly depends on all the remaining parameters of the model ---
the external magnetic field $h_{B}$, the external injection current
$\gamma $, and also the shape parameter $\sigma $. An analogous
statement is also valid for multifluxon distributions of the magnetic
flux in the junction, including unstable ones. The minimum length for
multifluxon distributions $\Phi ^{n}$ falls off rapidly with increasing
index $n$. For example, at otherwise equal parameters the vortex $\Phi
^{2}$ exists in Josephson junctions with $l>11.64$ (see Fig.\ \ref{lamin_ls07}).
Consequently, if $\sigma  = 0.07$, $h_{B} = 1$, and $\gamma  = 0$, then
at lengths $l<5.23$ the junction is in all respects short, and the
only stable distribution in the junction is the Meissner one. For
$5.23<l<11.64$ the junction is short for $\Phi ^{2}$, but a nontrivial
stable vortex $\Phi ^{1}$ can exist in it.

Let us now consider the question of constructing by numerical means the
critical current versus magnetic field relation, which is determined
implicitly by Eq.\ \eqref{gamhb} for each magnetic flux distribution in the
Josephson junction. The importance of this problem stems from the
possibility of measuring this relation experimentally.\cite{3,5,7,8,9,10}
We note that for different configurations of the magnetic flux in
the Josephson junction the values of the critical parameters --- in
particular, the critical current and magnetic field --- can be
substantially different. One should therefore identify the critical
parameters for specific distributions and for the Josephson junction as
a whole.

The curves of $\lambda_{\rm min}(h_{B})$ for the $M$ distribution and
the first few stable vortices in a Josephson junction at current $\gamma
 = 0$ and $\sigma  = 0$ are demonstrated in Fig.\ \ref{lam_hbs0}. For
comparison the analogous curves for $\sigma  = 0.07$ are shown in Fig.\ \ref{lamin_hb}. Each curve has two zeroes, the distance between which
determines the stability region of the vortex upon variation of the
magnetic field $h_{B}$. The zeroes themselves are critical values of
the field $h_{B}$ at zero current $\gamma $.

\myfigures{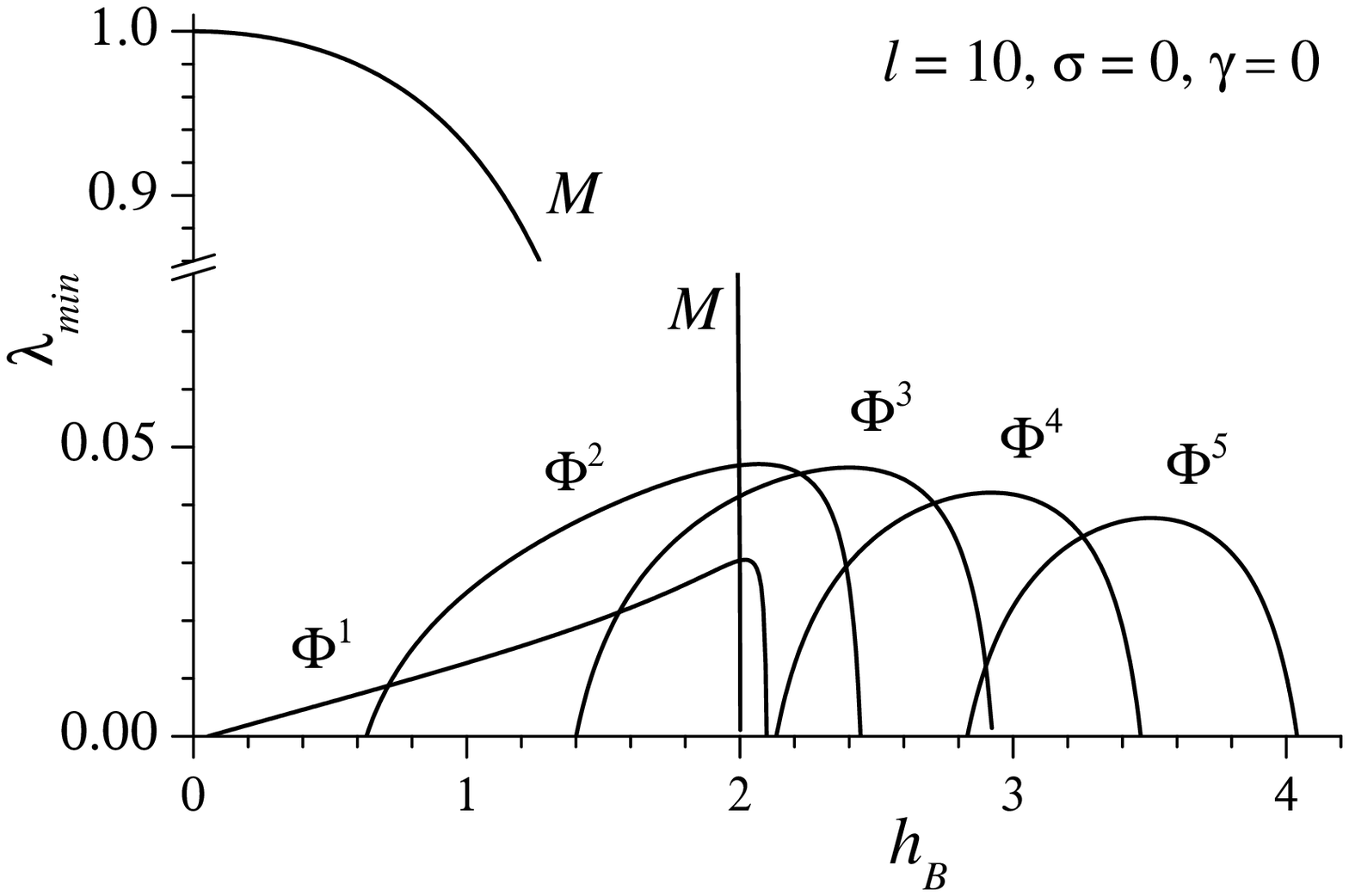}{0.44}{The $\lambda_{\rm min}(h_{B})$ curves for $\sigma = 0$.}{0.47} {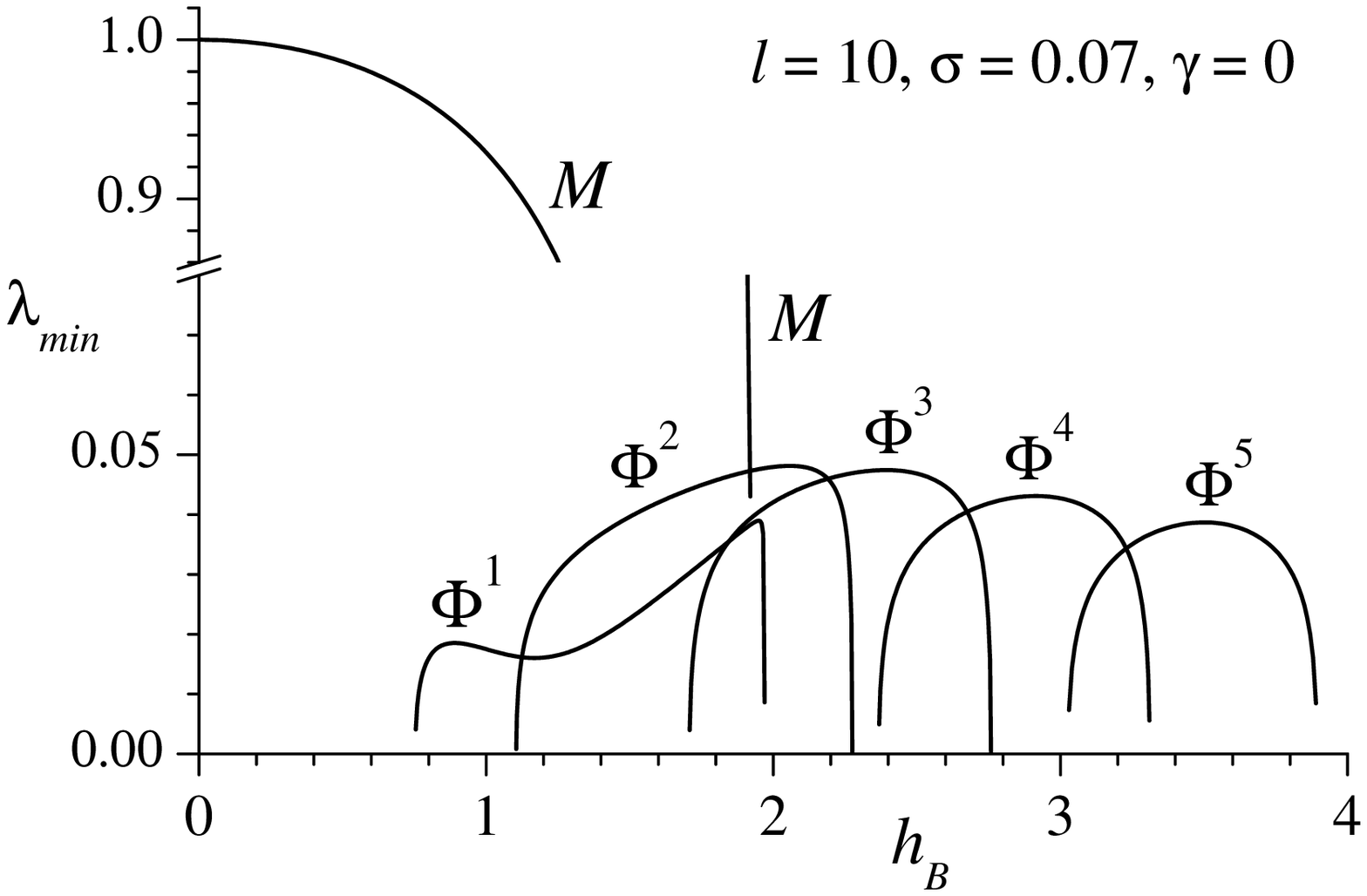}{0.44}{The $F(h_{B})$ curves for $\sigma  = 0.07$. }{0.51}

It is well seen that the role of the shape parameter $\sigma $ is most
important at small values of $h_{B}$. In particular, the $\Phi ^{1}$
vortex at $\sigma  = 0$ exists already at $h_{B}\approx 0.054$. At
$\sigma  = 0.07$, however, the existence region in respect to field
$h_{B}$ is considerably compressed, and the vortex exists starting at
$h_{B}>0.75$. The amount of compression of the $\lambda_{\rm
min}(h_{B})$ curves for different vortices decreases rapidly with
increasing $h_{B}$.

To complete the picture, Fig.\ \ref{fen_hbs07} shows graphs of the total
energy of several magnetic flux distributions in the Josephson junction
\begin{equation}\label{fen}
    F(p) = \int\limits_0^l \left[ \frac{1}{2}\,\varphi\,{'\,^2}  + \left(1 - \cos\varphi\right) \right]\,dx - h_B \Delta \varphi - l\gamma \varphi(0)\,,
\end{equation}
as a function of the external magnetic field $h_{B}$ for $\sigma = 0.07$ and current $\gamma  = 0$. The energy values are divided by the
total energy $F[\Phi ^{1}] = 8$ of an isolated fluxon in an infinite
Josephson junction.\cite{5} The solid and dashed curves show the energy
of the stable and unstable distributions, respectively. The points of
tangency of the solid and dashed curves are the points at which
the magnetic flux loses stability.

Let us now consider the situation for $\gamma \neq 0$. To find the
dependence of the critical current $\gamma_{c}$ on the external
magnetic field $h_{B}$ it is necessary to determine the stability
region with respect to current for the distributions $M$, $\Phi ^{1}$,
$\Phi ^{2}$, etc. The results of such calculations for certain values
of $h_{B}$ are given in Figs.\ \ref{lagam_hb16} --- \ \ref{lagam_f2}.

\myfigures{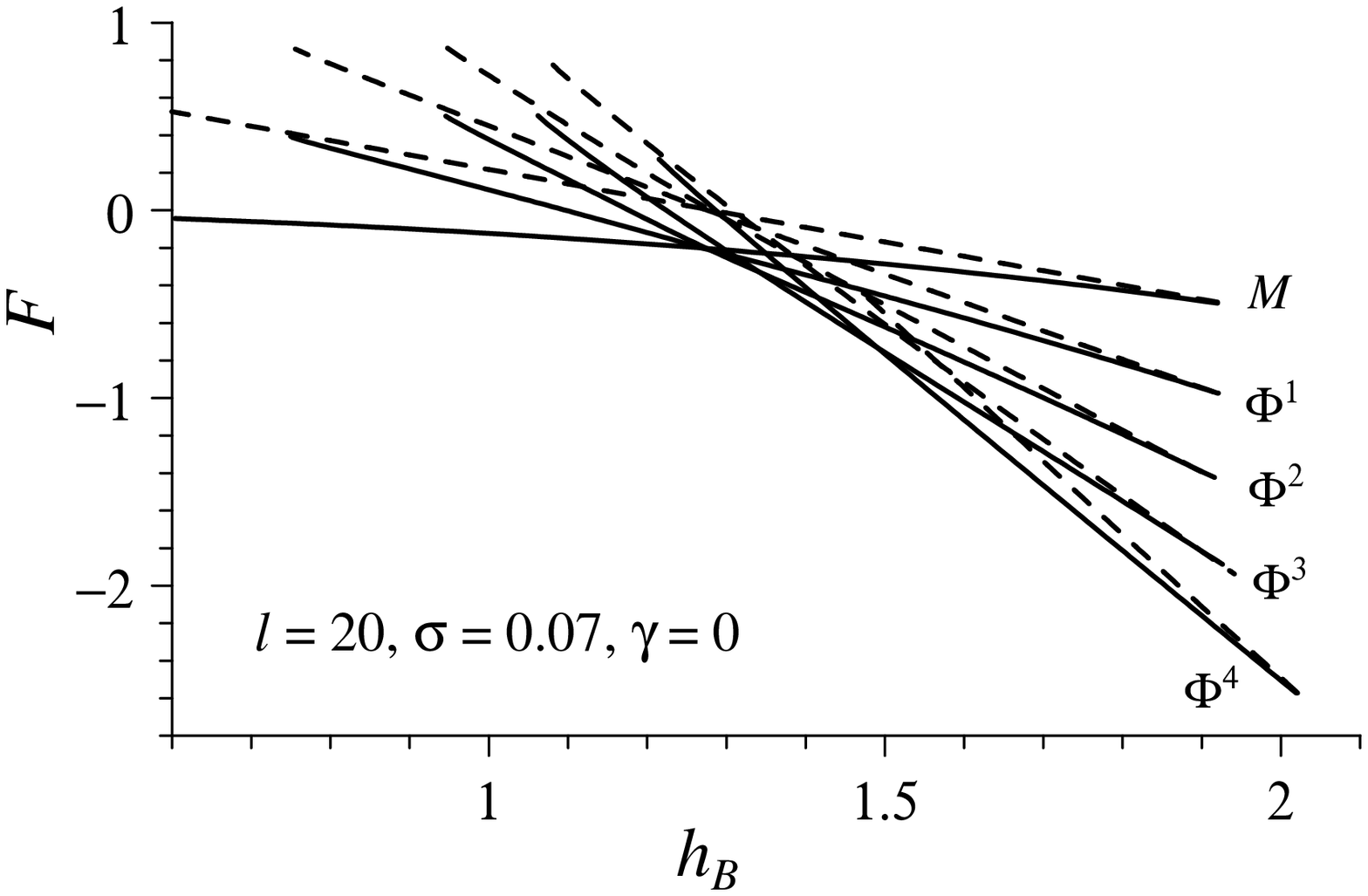}{0.44}{The $F(h_{B})$ curves for $\sigma  = 0.07$. }{0.49} {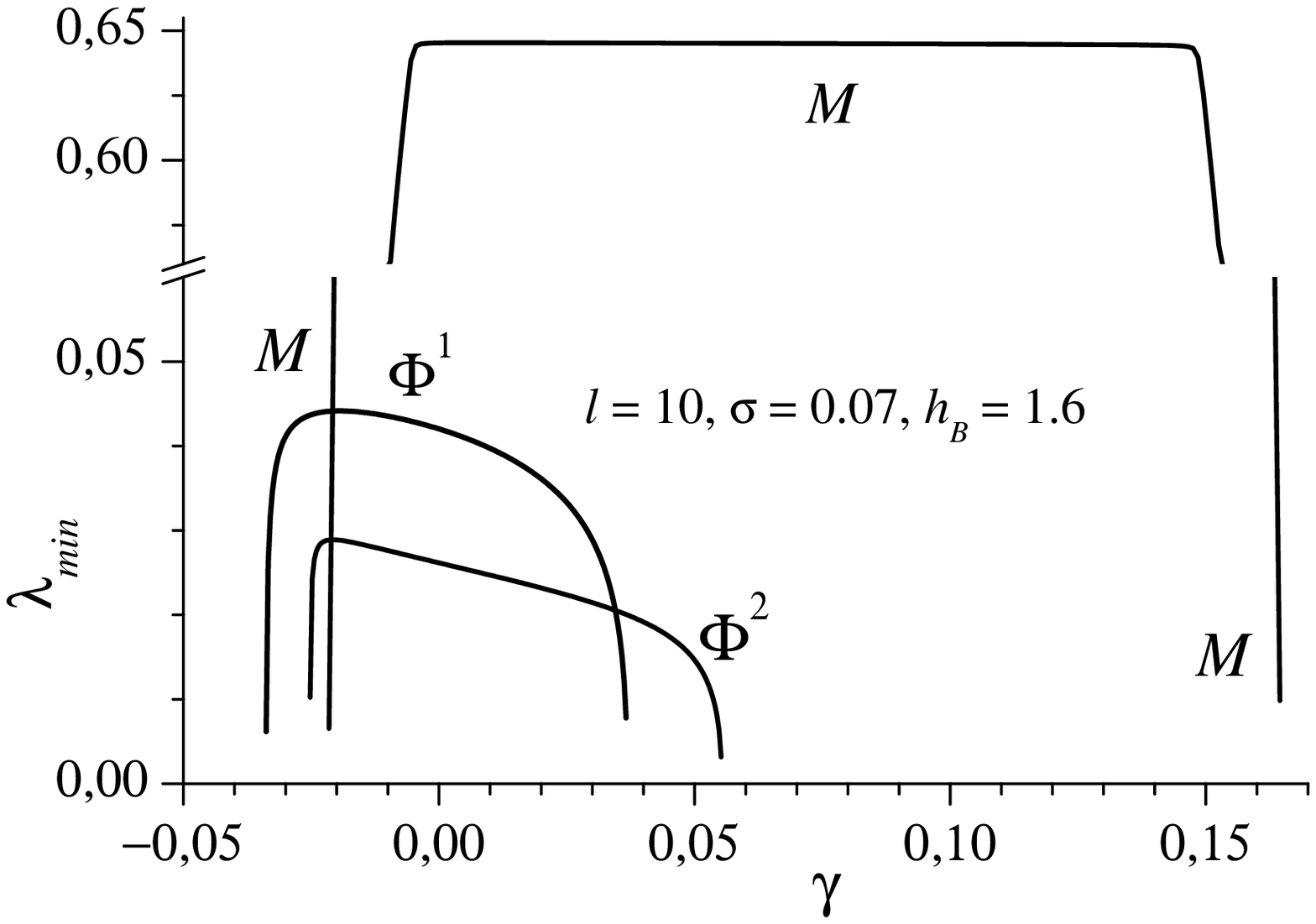}{0.42}{The $\lambda_{\rm min}(\gamma )$ curves for $h_{B} = 1.6$.
}{0.49}

Figure \ref{lagam_hb16} shows the $\lambda_{\rm min}(\gamma )$ curves for
stable solutions of the problem \eqref{stat} in a field $h_{B} = 1.6$. The
distances between zeroes of the functions are the stability intervals
of the corresponding distributions with respect to the current $\gamma
$. The right and left zeroes of the function $\lambda_{\rm min}(\gamma
)$ are the positive and negative critical currents, respectively, of
the distribution in the given field $h_{B}$. Because of the asymmetry
of the boundary conditions \eqref{bca} and \eqref{bcb} for $\gamma >0$ the critical
current of the Meissner distribution [which we denote by $\gamma
_{c}(M)$] is the highest, but for $\gamma <0$ the largest in modulus is
the critical current $\gamma_{c}(\Phi ^{1})$ of the vortex $\Phi
^{1}$. Consequently, in a field $h_{B} = 1.6$ the positive critical
current of the junction is $\gamma_{c} = \gamma_{c}(M)$, while the
negative critical current is $\gamma_{c} = \gamma_{c}(\Phi ^{1})$.

We note that for the junction geometry under consideration the curve for
the $M$ distribution has a characteristic plateau --- the ``breakoff''
of the Meissner solution sets in at a rather large modulus of the
external current.

\myfigures{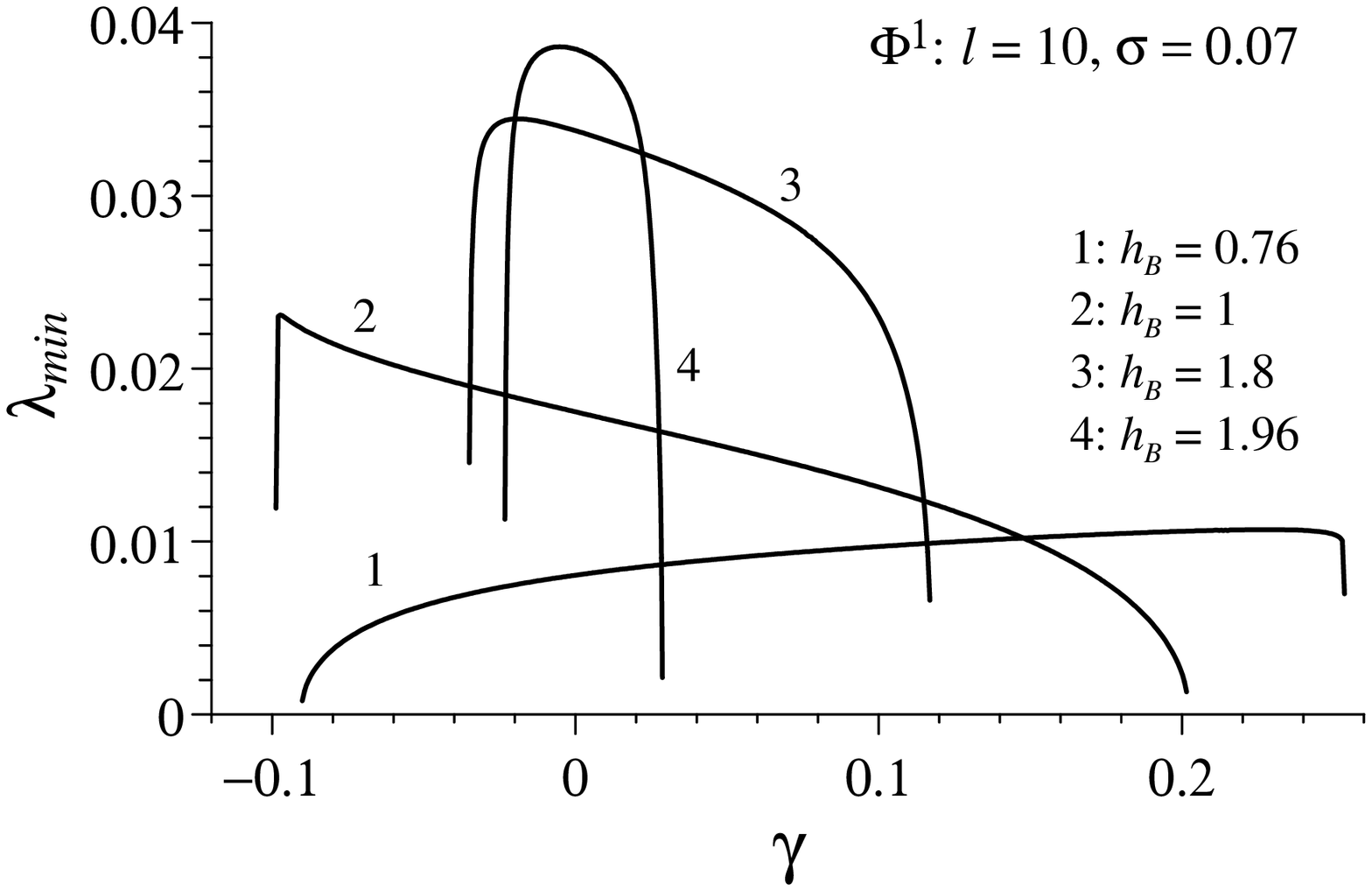}{0.45}{The $\lambda_{\rm min}(\gamma )$ curves for $\Phi ^{1}$. }{0.5} {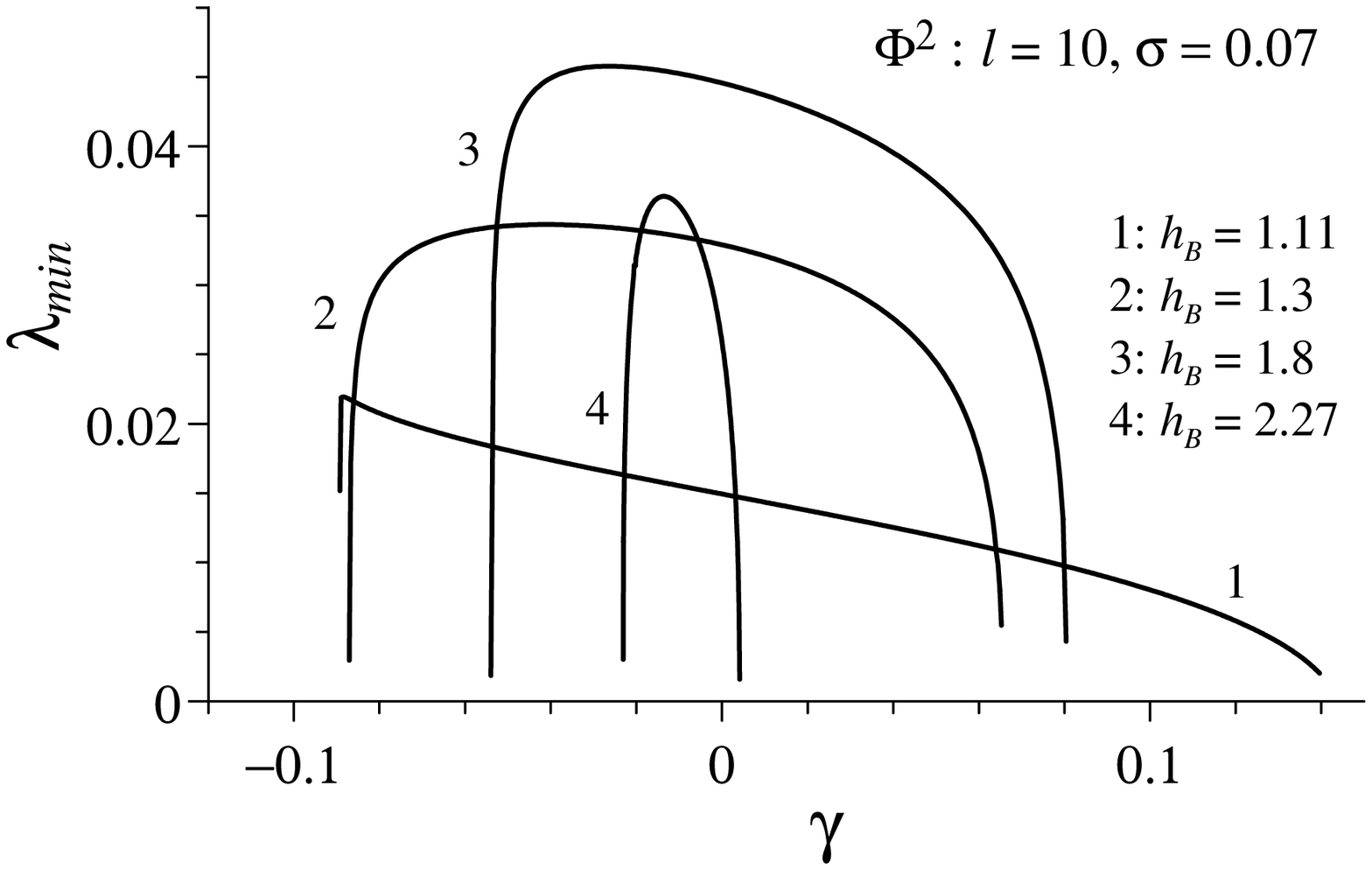}{0.45}{The $\lambda_{\rm min}(\gamma )$ curves for $\Phi ^{2}$. }{0.48}

Figure \ref{lagam_f} demonstrates the $\lambda_{\rm min}(\gamma )$ curves
for the main fluxon $\Phi ^{1}$ at $\sigma  = 0.07$ and several
values of $h_{B}$. The analogous curves for the $\Phi ^{2}$
distribution are shown in Fig.\ \ref{lagam_f2}. We note that with increasing
external magnetic field $h_{B}$ the stability region of the vortices
with respect to current is narrowed. Consequently, by varying $h_{B}$
one can construct the bifurcation curves for individual vortices to
acceptable accuracy and from them determine the critical values of the
current $\gamma_{c}$ for a Josephson junction.

A method of direct calculation of the bifurcation points of the
vortices in a Josephson junction was proposed in Refs.\ \cite{19}
and \cite{20}.

Figure \ref{critcur_f} shows the critical curves (1.8) for the main stable
fluxon $\Phi ^{1}$ for values of the shape parameter $\sigma  = 0$,
$\sigma  = 0.001$, and $\sigma  = 0.07$. The solid curves correspond to
a current $\gamma >0$ and the dashed curves to $\gamma <0$. We note
that at a value $h_{B}\approx 1.273$ the critical curves corresponding
to $\gamma >0$ intersect with each other and with the curve
corresponding to $\sigma  = 0$ in some narrow region. This means that
in that region of $h_{B}$ the critical current depends only slightly on
the shape of the junction. Geometrically the influence of $\sigma $
reduces to a rotation of the critical curves clockwise about the curve
for $\sigma  = 0$ by an angle that depends on $\sigma $. An analogous
effect takes place for the critical curves corresponding to $\gamma
<0$.

The critical curve $\gamma_{c}(h_{B})$ for a junction is constructed as
the envelope of the critical curves corresponding to different magnetic
flux distributions in the junction. In other words, the critical curve
consists of pieces of the critical curves for individual states with
the largest module of the critical current at a given $h_{B}$. The part
of the critical curve corresponding to the interval $h_{B}\in[0,2.8)$
is illustrated in Fig.\ \ref{critcur_hb}. For example, let $h_{B} = 1.4$. At a
current $\gamma  = 0$ there are five different magnetic field
distributions in the junction, which are described above (see Fig.\ \ref{allsol_hb14}). With increasing current $\gamma $ in the direction of
positive values the vortices lose stability in the following order:

$$\Phi^4 \to \Phi^3 \to \Phi^2  \to \Phi^1 \to M\,.$$

The last to break off is the Meissner distribution, the critical
current of which, $\gamma_{c}(M)\approx 0.156$, is the critical
current of the junction at a fixed value of the external field.
Consequently, the resistive regime in the junction at $h_{B} = 1.4$
exists for $\gamma >0.156$.

If the current $\gamma $ is increased from zero in the negative
direction, then the breakoff of the distributions occurs in the
opposite order: 
$$M\rightarrow \Phi ^{1}\rightarrow \Phi ^{2}\rightarrow \Phi ^{3}\rightarrow \Phi ^{4}\,,$$
and the critical current for the junction will be that of the vortex $\Phi ^{4}$: $\gamma_{c}(\Phi^{4})\approx -0.039$.
\myfigures{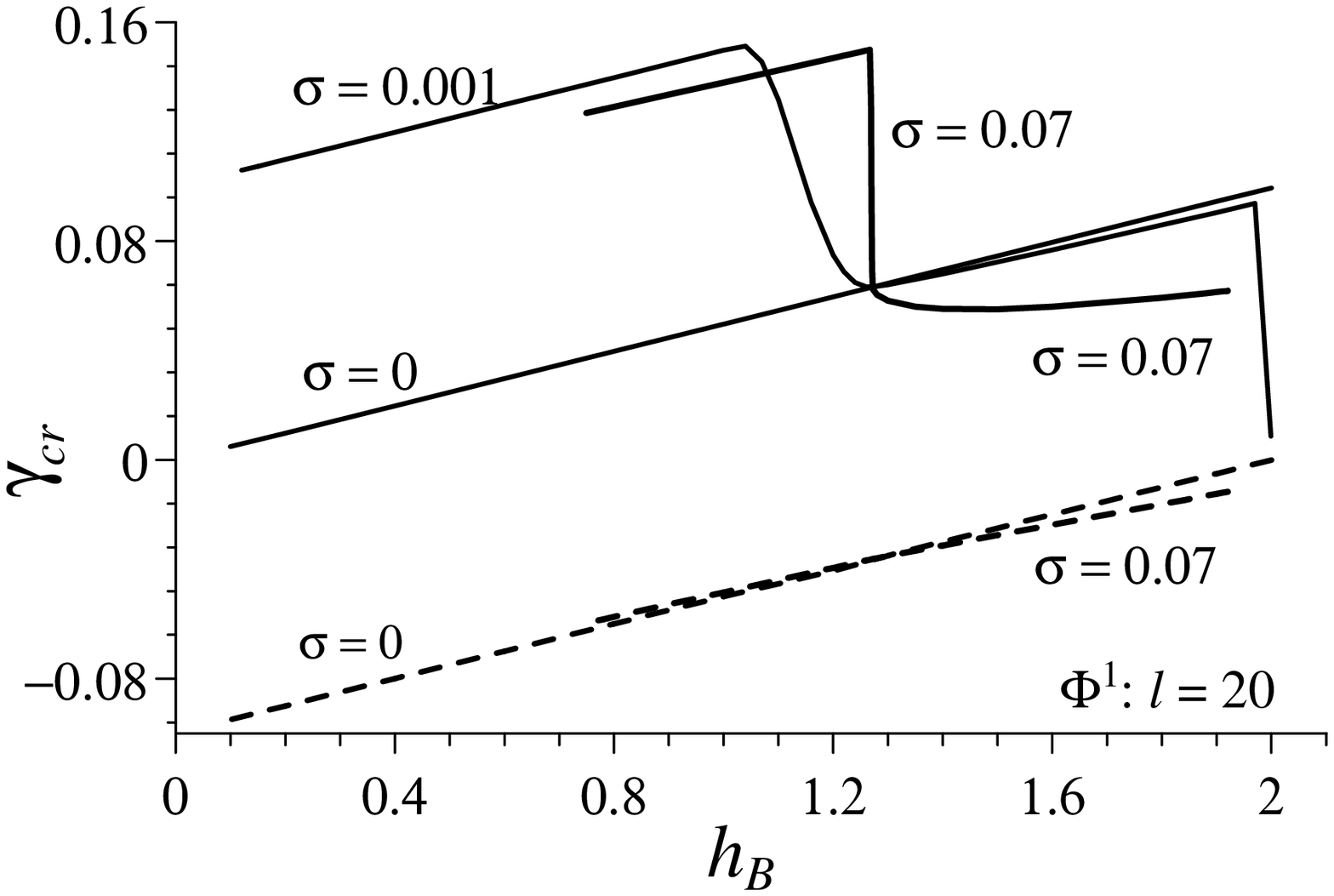}{0.46}{The critical curve of the fluxon $\Phi^{1}$.}{0.5} {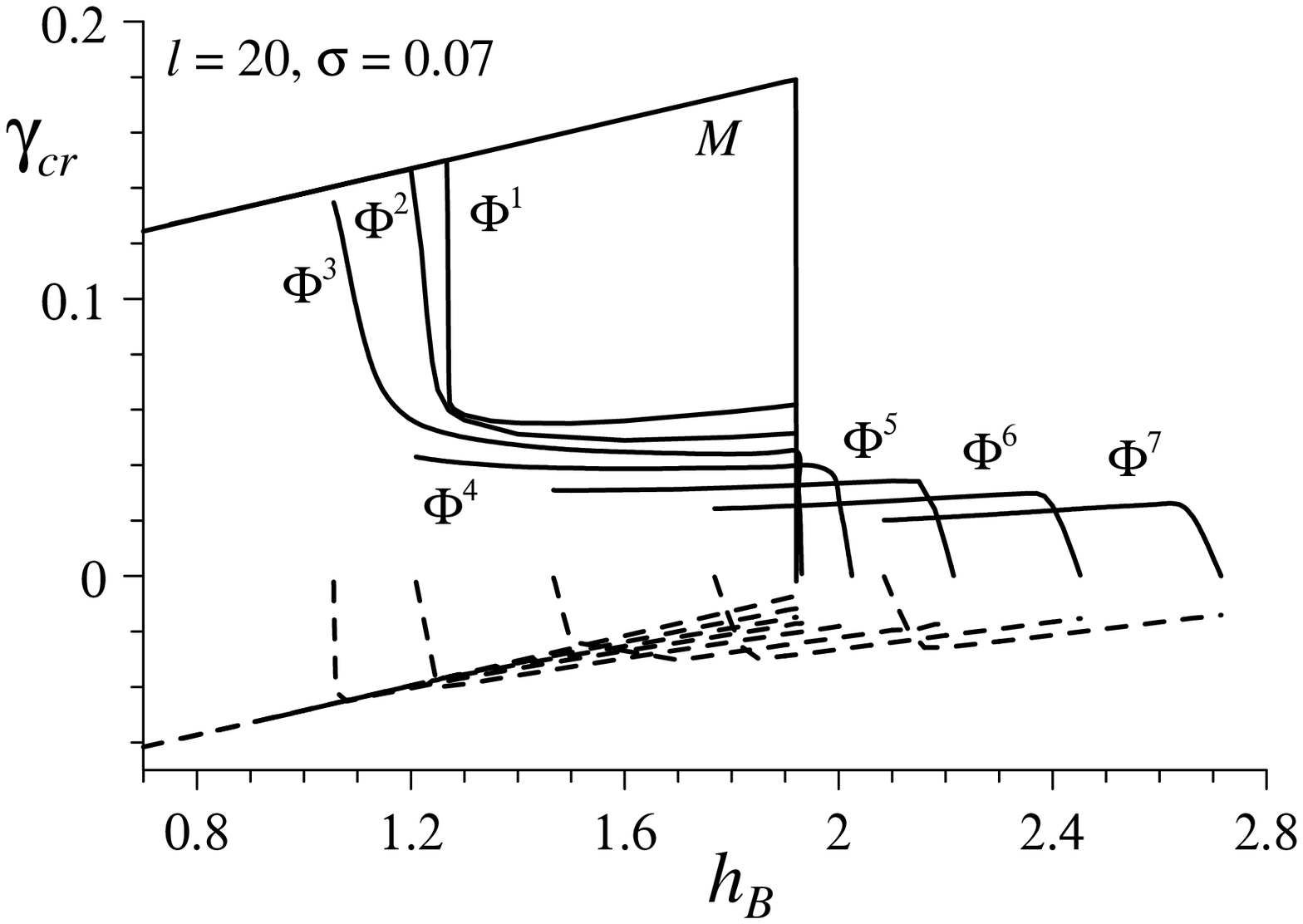}{0.44}{The critical curve of the junction.}{0.48}

Analogously, for example, at $h_{B} = 2.1$ the positive critical
current for the junction is that of the vortex $\Phi ^{5}$, $\gamma
_{c}(\Phi ^{5})\approx 0.034$, while the negative critical current is
that of the vortex $\Phi ^{6}$, $\gamma_{c}(\Phi ^{6})\approx 0.024$,
etc.

Of course, the ``switching'' of the vortices depends on all the
remaining parameters of the problem. For example, for a junction with $l= 10$ and $\sigma  = 0.07$ in an external field $h_{B} = 1.6$ the order of breakoff is as follows: $\Phi ^{1}\rightarrow \Phi ^{2}\rightarrow M$ if $\gamma >0$, and $M\rightarrow \Phi ^{2}\rightarrow \Phi ^{1}$ if
$\gamma <0$, as is clearly seen in Fig.\ \ref{lagam_hb16}.

We note that our numerically constructed critical curve of the Josephson junction (see
Fig.\ \ref{critcur_hb}) matches well with the theoretical and experimental
results presented in Figs.\ 6a and 7a of \cite{3}.

\section{Conclusions}

We have carried out a numerical simulation of the magnetic flux
distributions and their bifurcations upon variation of the model
parameters in a long Josephson junction of exponentially varying width.
For a stability analysis each distribution is placed in corres\-pondence
with a Sturm--Liouville problem with a potential determined by the
given distribution. The distributions and, hence, the spectrum of the
SL problem depend on the model parameters. It is shown that at certain
critical (bifurcation) values of one or several parameters the magnetic
flux in the junction can have a change of stability. For vortex
configurations in a Josephson junction there are narrow regions of
variation of the boundary magnetic field in which the critical current
of the vortices depends only slightly on the shape parameter. Corresponding
to the magnetic flux vortices are minimum lengths of the junction at which
the distributions maintain their stability. The critical current
versus magnetic field curves were constructed by a numerical method for
the first few stable magnetic field flux distributions in a Josephson
junction. The critical curve for the junction as a whole is the
envelope of the critical curves of the individual distributions.

\subsection*{Acknowledgements}

The authors thank Yu. A. Kolesnichenko (FTINT, Kharkov) for helpful discussions.



\begin{thebibliography}{99}

\bibitem{1} A. T. Filippov, T. Boyadjiev, Yu. S. Gal'pern, and I.
V. Puzynin, Comm. JINR E17-89-106, Dubna (1989).

\bibitem{2} A. Benabdallah, J. G. Caputo, and A. C. Scott, Phys.
Rev. B {\bf 54}, 16139 (1996).

\bibitem{3} G. Carapella, N. Martucciello, and G. Costabile,
cond-mat/0203055.

\bibitem{4} K. K. Licharev, {\it Dynamics of Josephson Junctions and
Circuits}, Gordon and Breach, New York (1986).

\bibitem{5} Yu. S. Gal'pern and A. T. Filippov, Zh. \'{E}ksp. Teor.
Fiz. {\bf 86}, 1527 (1984) [Sov. Phys. JETP {\bf 59}, 894 (1984)].

\bibitem{6} B. M. Levitan and I. S. Sargsyan, {\it Sturm--Liouville and Dirac
Operators} [in Russian], Nauka, Moscow (1988).

\bibitem{7} Jhy-Jiun Chang and C. H. Ho, Appl. Phys. Lett. {\bf
45}, 192 (1984).

\bibitem{8} A. N. Vystavkin, Yu. F. Drachevskii, V. P.
Koshelets, and I. L. Serpuchenko, Fiz. Nizk. Temp. {\bf 14}, 646
(1988) [Sov. J. Low Temp. Phys. {\bf 14}, 357 (1988)].

\bibitem{9} B. H. Larsen, J. Mygind, and A. V. Ustinov, Phys. Lett.
A {\bf 193}, 359 (1994).

\bibitem{10} B. H. Larsen, J. Mygind, and A. V. Ustinov, Physica B
{\bf 194--196}, 1729 (1994).

\bibitem{11} M. A. Krasnosel'ski\u{\i}, Usp. Matem. Nauk {\bf IX}, vyp. 3(61), 57 (1954).

\bibitem{12} J. B. Keller and S. Antman (ed.), {\it Bifurcation Theory and
Nonlinear Eigenvalue Problems}, Benjamin, Reading, Mass. (1969), Mir, Moscow (1974).

\bibitem{13} C. S. Owen and D. J. Scalapino, Phys. Rev. 164 (2), 538 (1967).

\bibitem{14} E. P. Zhidkov, G. I. Makarenko, and I. V. Puzynin, Fiz. \'{E}lem. Chastits At. Yadra {\bf 4}, 127 (1973) [Sov. J. Part. Nucl. {\bf 4}, 53 (1973)].

\bibitem{15} I. V. Puzynin, I. V. Amirkhanov, E. V. Zemlyanaya, V.
N. Pervushin, T. P. Puzynina, T. A. Strizh, and V. D. Lakhno,
Fiz. \'{E}lem. Chastits At. Yadra {\bf 30}, 97 (1999) [Phys. Part. Nucl.
{\bf 30}, 87 (1999)].

\bibitem{16} T. L. Boyadzhiev, Soobshchenie OIYaI R2-2002-101, Dubna (2002).

\bibitem{17} T. L. Boyadzhiev, D. V. Pavlov, and I. V. Puzynin,
Soobshchenie OIYaI R11-88-409, Dubna (1988).

\bibitem{18} T. L. Boyadzhiev, D. V. Pavlov, and I. V. Puzynin,
in {\it Numerical Methods and Applications}, Proc. Int. Conf. Num. Math.
and Appl., Bl. Sendov, R. Lazarov, and I. Dimov (eds.), Sofia, August 22--27, 1988.

\bibitem{19} D. W. McLaughlin and A. C. Scott, Phys. Rev. A {\bf
18}, 1652 (1978).

\bibitem{20} T. Boyadjiev and M. Todorov, Supercond. Sci.
Technology {\bf 14}, 1 (2002).


\end{thebibliography}
\end{document}